\documentclass[twocolumn]{aastex62}

\usepackage{CJK}
\usepackage{comment}
\usepackage{hyperref}

\usepackage{txfonts}
\usepackage{color,bm}

\newcommand{\eqref}[1]{(\ref{#1})}

\newcommand{\revise}[1]{#1}

\shorttitle{Transmission Spectra of Ringed Exoplanets}
\shortauthors{Ohno \& Fortney}

\begin{document}

\title{A Framework for Characterizing Transmission Spectra of Exoplanets with Circumplanetary Rings}

\author[0000-0003-3290-6758]{Kazumasa Ohno}
\affil{Department of Astronomy \& Astrophysics, University of California, Santa Cruz, 1156 High St, Santa Cruz, CA 95064, USA}

\author[0000-0002-9843-4354]{Jonathan J. Fortney}
\affil{Department of Astronomy \& Astrophysics, University of California, Santa Cruz, 1156 High St, Santa Cruz, CA 95064, USA}

\begin{abstract}
Recent observations revealed that several extremely low-density exoplanets show featureless transmission spectra.
While atmospheric aerosols are a promising explanation for both the low density and featureless spectra, there is another attractive possibility: the presence of circumplanetary rings.
Previous studies suggested that rings cause anomalously large transit radii. 
However, it remains poorly understood how rings affect the transmission spectrum. 
Here, we provide a framework to characterize the transmission spectra of ringed exoplanets.
We develop an analytical prescription to include rings in the transmission spectra for arbitrarily viewing geometries.
We also establish a simple post-processing model that can include the ring's effects on precomputed ring-free spectra.
The ring flattens the transmission spectrum for a wide range of viewing geometries, consistent with the featureless spectra of extremely low-density exoplanets.
Near-future observations by \emph{JWST} at longer wavelengths would be able to distinguish the aerosol and ring scenarios.
We also find that rocky rings might cause a silicate feature at $\sim10~{\rm {\mu}m}$ if the ring's optical depth is around unity. 
Thus, the ring's spectral features, if detected, would provide tight constrains on the physical properties of exoplanetary rings.
We also discuss the ring's stability and suggest that thick rings are sustainable only at the equilibrium temperature of $\la300~{\rm K}$ for the ring's age comparable to {\it Kepler} planets.
This might indicate the intrinsic deficit of thick rings in the {\it Kepler} samples, unless rings are much younger than the planets as suggested for Saturn.

\end{abstract}


\section{Introduction} \label{sec:intro}
Transmission spectroscopy is a powerful approach to explore exoplanetary atmospheres.
Observational efforts in the last decades have revealed a number of intriguing properties of exoplanetary atmospheres, such as the presence of gas-phase metals \citep[e.g.,][]{Charbonneau+02,Sedaghati+17,Nikolov+18}, the possible depletion of H$_2$O and CH$_4$ \citep[e.g.,][]{Madhusudhan+14,Benneke+19,Welbanks+19}, and prevalence of exotic clouds and hazes \citep[e.g.,][]{Pont+13,Kreidberg+14,Kreidberg+20,Knutson+14a,Sing+16}.
Recent studies have also started to probe the atmospheres of exoplanets orbiting near the habitable zone \citep[e.g.,][]{DeWit+16,DeWit+18,Benneke+19b,Tsiaras+19,Edwards+21,Gressier+21}.
The recently launched {\it James Webb Space Telescope} (JWST) would further revolutionize our understanding of the nature of exoplanets in the next decade.

Recent observations by the {\it Hubble Space Telescope} revealed that several exoplanets with extremely low bulk densities ($\rho_{\rm p}\la0.1~{\rm g~{cm}^{-3}}$) show featureless transmission spectra.
For example, \citet{Kreidberg+18} reported that the Neptune-mass planet WASP-107b with an extremely low-density \citep[$\rho_{\rm p}=0.134~{\rm g~{cm}^{-3}}$,][]{Piaulet+21} shows a NIR spectrum with a \revise{muted} water feature \revise{whose amplitude is} than the atmospheric scale height.
\citet{Libby-Roberts+20} also reported nearly flat NIR spectra for so-called the super-puff planets, namely, Kepler-51b ($\rho_{\rm p}=0.064~{\rm g~{cm}^{-3}}$) and -51d ($\rho_{\rm p}=0.038~{\rm g~{cm}^{-3}}$). 
\citet{Chachan+20} similarly reported featureless spectrum for Kepler-79d ($\rho_{\rm p}=0.08~{\rm g~{cm}^{-3}}$).
A recent statistical study by \citet{Dymont+21} showed that the water feature amplitude tends to be smaller for exoplanets with lower density.
The featureless spectra of extremely low-density exoplanets poses a puzzle, as the large atmospheric scale height should enhance the spectral features, in theory.

Several ideas have been proposed to explain the peculiar low density of those exoplanets.
One promising hypothesis is the presence of atmospheric aerosols.
\citet{Wang&Dai19} proposed that the extremely low-density planets undergo intense atmospheric escape entraining aerosols, which can enlarge the transit radius.
Several studies with aerosol microphysical models also showed that photochemical hazes under atmospheric escape can enlarge the transit radius significantly \citep{Gao&Zhang20,Ohno&Tanaka21}.
These studies also showed that the hazes can efficiently flatten the transmission spectra \citep[see also][]{Kawashima+19} in addition to explaining the extremely low planetary density.

Besides the hazy atmosphere hypothesis, there is another attractive possibility for the origin of extremely low-density planets: the presence of circumplanetary rings.
Circumplanetary rings are commonly present around solar system giant planets, namely, Jupiter, Saturn, Neptune, and Uranus \citep[for review, see, e,g.,][]{Esposito02,Tiscareno13}, as well as small dwarf planets, namely, (10199) Chariklo \citep{Brag-Ribas14}, (2060) Chiron \citep{Ortiz+15}, and (136108) Haumea \citep{Ortiz+17}.
\citet{Barnes&Fortney04} and \citet{Zuluaga+15} pointed out that the presence of a ring could potentially cause anomalously large transit radii for exoplanets because the ring causes additional transit occultation.
Later, \citet{Piro&Vissapragada19} discussed that several super-puffs could be explained by the presence of rings if the rings have a sufficiently large obliquity.
Planetary rings of great interest for a variety of astrophysical topics, such as satellite formation \citep{Charnoz+10,Crida&Charnoz12,Hyodo+15} and orbital dynamics \citep{Nakajima+19,Nakajima+20}, and inference of planetary interiors from ring seismology \citep{French+88,Marley91,Mankovich&Fuller21}.

Several previous studies examined the observable signatures of exoplanetary rings.
\citet{Barnes&Fortney04} showed that the ring leaves characteristic signatures in ingress and egress of transit light curves, which is detectable with a photometric precision of $\sim100~{\rm ppm}$ if the planet has a Saturn-like ring.
They also suggested that individual ring particles cause the forward-scattering of starlight, which may be observed as the excess stellar flux right before and after the transit ingress and egress, respectively.
\citet{Ohta+09} showed  that rings leave signatures in the radial velocity anomaly caused by the Rossiter-McLaughlin effect, allowing us to complementary check for the presence of a ring.
\citet{Zuluaga+15} suggested that rings cause a discrepancy of the observed stellar density from a true value, the so-called photo-ring effect, as rings affect the transit depth and duration time that are used to estimate a stellar radius to semi-major axis ratio associated with a stellar density. 
\citet{deMooij+17} further showed that rings cause a distortion of stellar line profiles, which also helps to constrain the ring physical properties.
\citet{Santos+15} suggested that rings enhance the observable reflected light from exoplanets \citep[see also][]{Dyudina+05}.
\citet{Heising+15} and \citet{Aizawa+17,Aizawa+18} performed systematic searches for ring signatures in transit light curves of {\it Kepler} planets; however, ringed exoplanets have not been discovered thus far.

Recently, \citet{Alam+22}, a companion paper of this study, reported a featureless NIR spectrum of an extremely low-density planet HIP41378 f observed by the {\it Hubble Space Telescope}.
HIP41378 f has an orbital period of $542$ days (equilibrium temperature of $294~{\rm K}$ for zero Bond albedo \revise{and full heat redistribution}) and a mass of $12M_{\rm \oplus}$, yielding a planetary density of $0.09~{\rm g~{cm}^{-3}}$ \citep{Santern+19}.
\citet{Akinsanmi+20} performed a Bayesian model comparison on the light curve of HIP41378 f and found that the \revise{Bayesian} evidence for the ringed planet scenario is comparable to the ring-free scenario.
\citet{Ohno&Tanaka21} suggested that the anomalously large radii of massive ($\ga10M_{\rm \oplus}$) planets are unlikely explained by photochemical hazes alone owing to the atmospheric scale height being much smaller than the planetary radius.
This places HIP41378 f on an intriguing target to search for the first exoplanetary ring.

It has been poorly understood how the ring affects the observable transmission spectra since most of the conventional models commonly assume ring-free planets.
\citet{Ohno&Tanaka21} introduced the a model transmission spectrum of a Kepler-51b like planet with a circumplanetary ring.
They showed that the ring acts to flatten the spectrum while weak spectral features remain.
However, they only considered a face-on ring, which would be unrealistic for most of the ringed exoplanets.
In this study, we present a generalized framework to characterize the atmospheric transmission spectra of ringed exoplanets with arbitrarily viewing geometries.
We also discuss the possible physical properties of exoplanetary rings and the effects of non-gray ring opacity.

The organization of this paper is as follows.
In Section \ref{sec:method}, we introduce how to include the effects of the ring in transmission spectrum models.
We also present a simple postprocessing method that can efficiently include the ring effects in precomputed ring-free spectra.
In Section \ref{sec:ring_property}, we estimate the possible surface density and particle size distribution of exoplanetary rings from a dynamical argument.
We also discuss the critical planetary equilibrium temperature above which a thick ring may be unstable.
In Section \ref{sec:results}, we show how ring properties, such as the inner edge radius and obliquity, affect the observable transmission spectrum.
We also discuss whether the spectral feature of the ring itself is observable or not.
In Section \ref{sec:discussion}, we discuss the implications for future observations, model caveats, and possible model extensions.
Finally, Section \ref{sec:summary} summarizes the paper.

\section{Transmission spectra of exoplanets with circumplanetary ring}\label{sec:method}
\begin{figure*}[t]
\centering
\includegraphics[clip, width=\hsize]{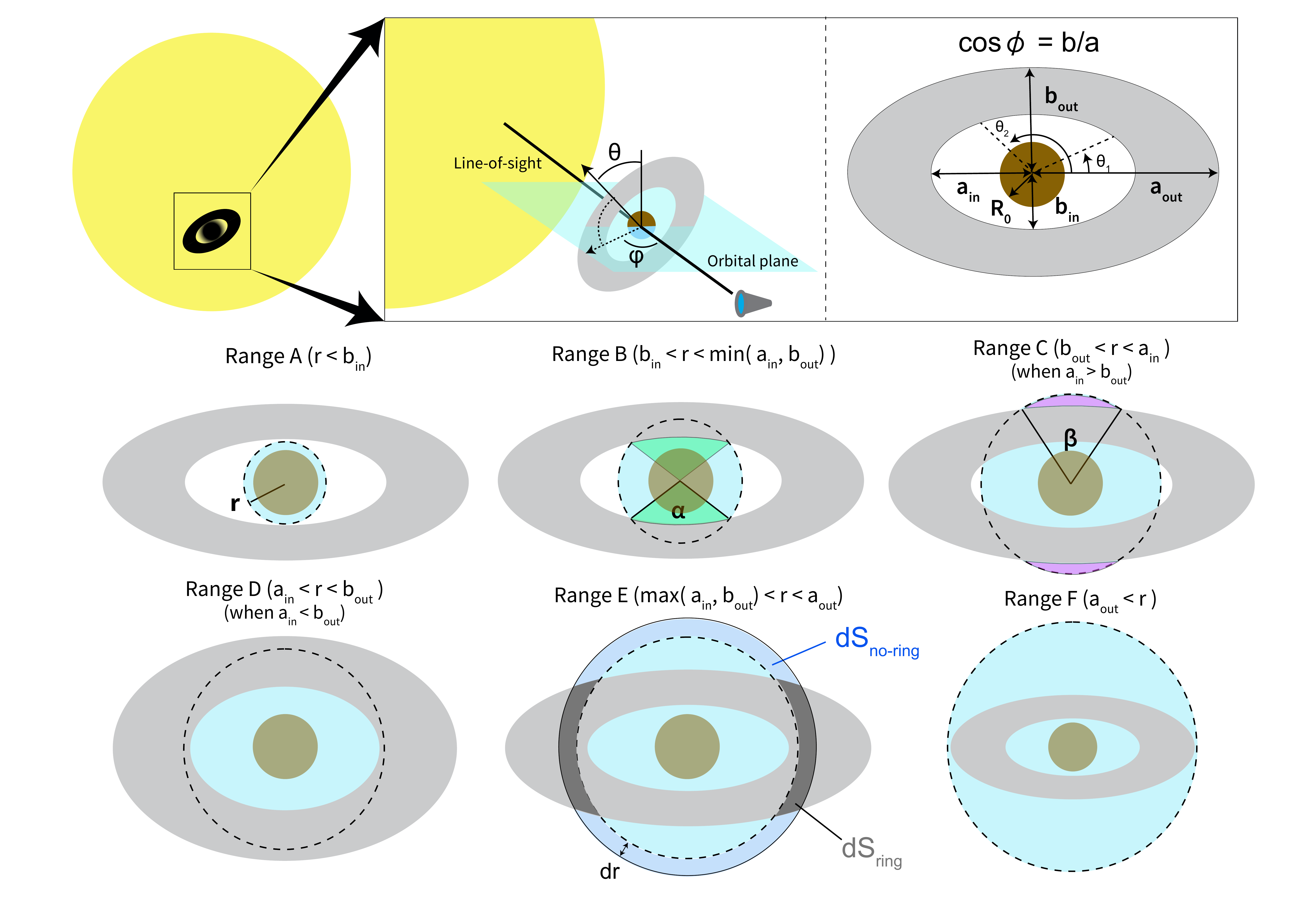}
\caption{Schematic illustration of possible viewing geometries of a ringed transit planet from an observer's perspective.
The ring appear as a projection onto the stellar disk, which is described by a concentric ellipse with semi-major and -minor axis lengths of $a_{\rm in/out}$ and $b_{\rm in/out}$ for inner/outer edge.
The gray regions express the projected ring, and the skyblue regions express the ring-free area within a circle with a radius of $r$, denoted by the black dashed lines.
\revise{One has to integrate the transmittance of each atmospheric annulus, which is expressed by the area between dashed and solid lines in the range E, over all radial distances.}
For computing the transmission spectrum of a ringed planet, one has to split each integrated annulus into the no-ring area $dS_{\rm no-ring}$ and ring-overlapped area $dS_{\rm ring}$, as denoted by the thick gray and blue regions in the range E.
\revise{We will use the areas of the green and purple shaded regions to derive the no-ring areas for range B, C, and E in Sections \ref{sec:range_B}, \ref{sec:range_C}, and \ref{sec:range_E}.}
The radial distance can be demarcated into six ranges according to the formula of $dS_{\rm no-ring}/dr$ and $dS_{\rm ring}/dr$, as summarized in the lower six graphics, above.
For further details, see Section \ref{sec:method}.
}
\label{fig:cartoon_ring}
\end{figure*}

In general, the transmission spectrum is computed as a transit radius at multiple wavelengths.
The spectrum can be formally computed from the surface integral over the stellar disk $S_{\rm disk}$, written by
\begin{equation}\label{eq:trans_ring}
    D = \frac{1}{\pi R_{\rm s}^2} \int_{S_{\rm disk}} [1-\exp{(-\tau)}]dS,
\end{equation}
where $\tau$ is the line-of-sight optical depth of the planet, including an atmospheric transmittance, and $R_{\rm s}$ is the stellar radius.
For planets without rings, assuming $\tau\rightarrow\infty$ at $r<R_{\rm 0}$, \revise{azimuthal} symmetry, and a planet being centered on the stellar disk, we obtain a conventional formula of the ring-free spectrum:
\begin{equation}\label{eq:spectrum_ring_free}
    D = \frac{1}{\pi R_{\rm s}^2}\left[\pi R_{\rm 0}^2+ \int_{\rm R_{\rm 0}}^{R_{\rm s}} 2\pi r[1-\exp{(-\tau)}]dr \right],
\end{equation}
\revise{where $r$ is the radial distance from the center of the planet}. 
If the planet has a ring, we have to split the integrated annulus into ring-overlapped and no-ring areas (see Figure \ref{fig:cartoon_ring}).
Then, Equation \eqref{eq:trans_ring} yields
\begin{eqnarray}\label{eq:spectrum_ring}
    D = \frac{1}{\pi R_{\rm s}^2}[&&\pi R_{\rm 0}^2+ \int_{\rm R_{\rm 0}}^{R_{\rm s}} [1-\exp{(-\tau)}]\frac{dS_{\rm no-ring}}{dr}dr\\
    \nonumber
    &&+ \int_{\rm R_{\rm 0}}^{R_{\rm s}}[1-\exp{[-(\tau+\tau_{\rm ring,LOS})]}]\frac{dS_{\rm ring}}{dr}dr ],
\end{eqnarray}
where $\tau_{\rm ring,LOS}$ is the line-of-sight optical depth of the circumplanetary ring, $S_{\rm ring}$ and $S_{\rm no-ring}$ are ring-overlapped and no-ring areas within a circle with a radius of $r$.
Since $S_{\rm ring}+S_{\rm no-ring}=\pi r^2$, the derivatives of the two areas satisfy the following relation
\begin{equation}\label{eq:relation}
    \frac{dS_{\rm ring}}{dr}+\frac{dS_{\rm no-ring}}{dr}=2\pi r
\end{equation}
Our task is to derive an explicit formula of either $dS_{\rm ring}/dr$ or $dS_{\rm no-ring}/dr$.
The expression of these derivative areas depends on the ring morphology, such as inner and outer edge radii, and the ring obliquity.

In the following subsections, we describe the no-ring area and its derivative at a radial distance split into six ranges  (see Figure \ref{fig:cartoon_ring}).
A similar geometrical argument can also be found in \citet{Piro18}.
We consider a circumplanetary ring with inner and outer edge radii of $R_{\rm in}$ and $R_{\rm out}$.
The ring has its own obliquity $\theta$, the angle between the orbital normal and rotation axis.
We also define the angle $\varphi$ between a line-of-sight line and the rotation axis projected onto the orbital plane.
For example, the observer views the ring from an edge-on perspective for $\varphi=\pi/2$ regardless of its obliquity.
When the ring has a vertical optical depth of $\tau_{\rm ring}$, the line-of-sight optical depth is given by $\tau_{\rm ring,LOS}=\tau_{\rm ring}/\sin{(\theta)}\cos{(\varphi)}$.
For the transit observation, the ring is observed as a projection onto the stellar disk, which can be expressed by a concentric ellipse centered on the planet (see Figure \ref{fig:cartoon_ring}).
For the latter convenience, we introduce the major and minor axes of the inner/outer edges of the projected ring as $a_{\rm in/out}$ and $b_{\rm in/out}$.
\revise{Assuming the observer's line-of-sight aligned with the orbital plane, the major and minor axes are related to the ring's obliquity $\theta$ and $\varphi$ as $a_{\rm in/out}=R_{\rm in/out}\max{[\cos{(\varphi)},\sin{(\theta)]}}$ and $b_{\rm in/out}=R_{\rm in/out}\min{[\cos{(\varphi)},\sin{(\theta)]}}$.}
If one defines the angle between the ring and a sky-plane $\phi$ as in \citet{Akinsanmi+20} (their angle $i_{\rm r}$), the line-of-sight optical depth is given by $\tau_{\rm ring,LOS}=\tau_{\rm ring}/\cos{\phi}$, and the major and minor axes are instead written by $a_{\rm in/out}=R_{\rm in/out}$ and $b_{\rm in/out}=R_{\rm in/out}\cos{\phi}$.
We note that the above definition of $\phi$ implicitly assumes $\varphi=0$.

\subsection{Range A: $r<b_{\rm in}$}
The ring does not affect the line-of-sight optical depth for $r<b_{\rm in}$, as shown in the first of six graphics in Figure \ref{fig:cartoon_ring}.
The no-ring area is trivially given by 
\begin{equation}\label{eq:S_A}
S_{\rm no-ring}=\pi r^2.
\end{equation}
\revise{We note that the no-ring area involves the optically thick solid part of the planet, i.e., $\pi R_{\rm 0}^2$.
Since the area of the solid part is independent of the radial distance $r$, it does not affect our subsequent derivation.}
The derivative of the no-ring area is thus given by
\begin{equation}\label{eq:dS_A/dr}
 \frac{dS_{\rm no-ring,A}}{dr}=2\pi r.
\end{equation}

\subsection{Range B: $b_{\rm in}<r<\min{(b_{\rm out},a_{\rm in})}$}\label{sec:range_B}
At the radial distance of $b_{\rm in}<r<\min{(b_{\rm out},a_{\rm in})}$, a part of the atmosphere is overlapped by the projected ring (see the second of six graphics in Figure \ref{fig:cartoon_ring}).
In general, the area of an ellipsoidal sector from angle $\theta_{\rm 1}$ to $\theta_{\rm 2}$, $S(\theta_{\rm 1},\theta_{\rm 2})$, is given by
\begin{eqnarray}\label{eq:area_fan}
    \nonumber
    S(\theta_{\rm 1},\theta_{\rm 2})&=&\int_{\rm \theta_{\rm 1}}^{\theta_{\rm 2}}\frac{1}{2}\frac{a^2}{1+(a^2/b^2-1)\sin^2{\theta}}d\theta\\
    &=&\frac{1}{2}ab \left[\tan^{-1}{\left( \frac{a}{b}\tan{\theta_{\rm 2}}\right)} - \tan^{-1}{\left( \frac{a}{b}\tan{\theta_{\rm 1}}\right)}\right],
\end{eqnarray}
where $a$ and $b$ are the major and minor axes of an ellipse.
As seen in Figure \ref{fig:cartoon_ring}, the no-ring area consists of two circular sectors and ellipsoidal sectors.
Using Equation \eqref{eq:area_fan}, the area of the ellipsoidal sector (a green shaded part in Figure \ref{fig:cartoon_ring}) is given by \footnote{We note that \citet{Piro18} derived the area of the same ellipsoidal sector (a blue sector in their Figure 2, corresponding to a green sector in our Figure \ref{fig:cartoon_ring}) as $ab\alpha/2$ using Cavallieri's principle, which is different from our Equation \eqref{eq:ellip_sector}. We find that this difference is due to their incorrect use of  Cavallieri's principle.  Cavallieri's principle yields the area of an ellipsoidal sector by multiplying $b/a$ to the area of circular sector $a^{2}x/2$, where $x$ is the opening angle of the corresponding {\it circular sector}, not the opening angle of the ellipsoidal sector $\alpha$. When the sector is centered on the minor axis as in Figure \ref{fig:cartoon_ring}, the angle $x$ is related to $\alpha$ as $a/b=\tan{(\pi/2-x/2)}/\tan{(\pi/2-\alpha/2)}$, which yields $x=\pi-2\tan^{-1}{(a/b\tan{(\alpha/2)})}$. Thus, the ellipsoidal sector area $abx/2$ derived by Cavallieri's principle, when properly implemented, is equivalent to Equation \eqref{eq:ellip_sector}.}
\begin{equation}\label{eq:ellip_sector}
    S\left(\frac{\pi-\alpha}{2},\frac{\pi+\alpha}{2}\right)=\frac{ab}{2}\left[ \pi-2\tan^{-1}{\left( \frac{a}{b\tan{(\alpha/2)}}\right)} \right],
\end{equation}
where the $\alpha$ is the opening angle defined by 
\begin{equation}\label{eq:angle_alpha}
    \alpha=2\sin^{-1}{\left( \sqrt{\frac{1-(b_{\rm in}/r)^2}{1-(b_{\rm in}/a_{\rm in})^2}}\right)}.
\end{equation}
Using Equation \eqref{eq:ellip_sector}, the no-ring area is given by
\begin{equation}\label{eq:S_B}
    S_{\rm no-ring,B}=r^2(\pi-\alpha)+a_{\rm in}b_{\rm in}\left[ \pi- 2\tan^{-1}{\left( \frac{a_{\rm in}}{b_{\rm in}\tan{(\alpha/2)}}\right)} \right],
\end{equation}
The derivative of the no-ring area is then given by
\begin{equation}\label{eq:dS_B/dr}
    \frac{dS_{\rm no-ring,B}}{dr}=2r(\pi-\alpha) - \left(r^2-\frac{a_{\rm in}^2b_{\rm in}^2}{a_{\rm in}^2\cos^2{(\alpha/2)}+b_{\rm in}^2\sin^2{(\alpha/2)}}\right)\frac{d\alpha}{dr},
\end{equation}
where the derivative of the opening angle is given by
\begin{equation}\label{eq:dalpha}
    \frac{d\alpha}{dr}=\frac{2b_{\rm in}}{r^2}\left[ \left(1-\frac{b_{\rm in}^2}{r^2}\right)\left(1-\frac{r^2}{a_{\rm in}^2}\right)\right]^{-1/2}.
\end{equation}

\subsection{Range C: $b_{\rm out}<r<a_{\rm in}$ when $a_{\rm in}>b_{\rm out}$}\label{sec:range_C}
For the ring with low obliquity and near face-on viewing configuration, the outer edge of the projected ring can have a minor axis shorter than the major axis of the projected inner edge (see the third of six graphics in Figure \ref{fig:cartoon_ring}). 
In this case, the circle has extra no-ring area\revise{s} outside the projected outer edge (purple shaded part\revise{s} in Figure \ref{fig:cartoon_ring}).
The extra no-ring area can be computed by subtracting an ellipsoidal sector from a circular sector with an opening angle of $\beta$, defined by
\begin{equation}\label{eq:angle_beta}
    \beta=2\sin^{-1}{\left( \sqrt{\frac{1-(b_{\rm out}/r)^2}{1-(b_{\rm out}/a_{\rm out})^2}}\right)}.
\end{equation}
Utilizing Equation \eqref{eq:ellip_sector}, the no-ring area $S_{\rm C}$ is given by
\begin{equation}\label{eq:S_C}
    S_{\rm no-ring,C}=S_{\rm no-ring,B}+ r^2 \beta-a_{\rm out}b_{\rm out}\left[ \pi- 2\tan^{-1}{\left( \frac{a_{\rm out}}{b_{\rm out}\tan{(\beta/2)}}\right)} \right].
\end{equation}
The derivative of the no-ring area is expressed by
\begin{eqnarray}\label{eq:dS_C/dr}
    \frac{dS_{\rm no-ring,C}}{dr}&=&\frac{dS_{\rm no-ring,B}}{dr}+2\beta r\\
    \nonumber
    && + \left(r^2-\frac{a_{\rm out}^2b_{\rm out}^2}{a_{\rm out}^2\cos^2{(\beta/2)}+b_{\rm out}^2\sin^2{(\beta/2)}}\right)\frac{d\beta}{dr},
\end{eqnarray}
where the derivative of the opening angle is given as the same as Equation \eqref{eq:dalpha}:
\begin{equation}\label{eq:dbeta}
    \frac{d\beta}{dr}=\frac{2b_{\rm out}}{r^2}\left[ \left(1-\frac{b_{\rm out}^2}{r^2}\right)\left(1-\frac{r^2}{a_{\rm out}^2}\right)\right]^{-1/2}.
\end{equation}

\subsection{Range D: $a_{\rm in}<r<b_{\rm out}$ when $a_{\rm in}<b_{\rm out}$}
For a nearly face-on ring with high obliquity, the outer edge of the projected ring may have a minor axis longer than the major axis of the projected inner edge (see the fourth of six graphics in Figure \ref{fig:cartoon_ring}). 
In this context, the whole projected inner hole is embedded within a circle with a radius of $r$.
The no-ring area is thus given by
\begin{equation}\label{eq:S_D}
S_{\rm no-ring,D}=\pi a_{\rm in}b_{\rm in}. 
\end{equation}
Thus, the derivative is trivially given by
\begin{equation}\label{eq:dS_D/dr}
 \frac{dS_{\rm no-ring,D}}{dr}=0.
\end{equation}

\subsection{Range E: $\max{(a_{\rm in},b_{\rm out})}<r<a_{\rm out}$}\label{sec:range_E}
At the radial distance of $\max{(a_{\rm in},b_{\rm out})}<r<a_{\rm out}$, the no-ring area consists of a whole inner hole and the area outside the projected outer edge (see the fifth of six graphics in Figure \ref{fig:cartoon_ring}).
Replacing $S_{\rm no-ring,B}$ by $\pi a_{\rm in}b_{\rm in}$ in Equation \eqref{eq:S_C}, we obtain the no-ring area as 
\begin{equation}\label{eq:S_E}
    S_{\rm no-ring,E}=\pi a_{\rm in}b_{\rm in}+ r^2 \beta-a_{\rm out}b_{\rm out}\left[ \pi- 2\tan^{-1}{\left( \frac{a_{\rm out}}{b_{\rm out}\tan{(\beta/2)}}\right)} \right].
\end{equation}
Its derivative is given by
\begin{equation}\label{eq:dS_E/dr}
    \frac{dS_{\rm no-ring,E}}{dr}=2\beta r + \left(r^2-\frac{a_{\rm out}^2b_{\rm out}^2}{a_{\rm out}^2\cos^2{(\beta/2)}+b_{\rm out}^2\sin^2{(\beta/2)}}\right)\frac{d\beta}{dr}.
\end{equation}

\subsection{Range F: $a_{\rm out}<r$}
The whole projected ring is within the circle at the radial distance of $r>a_{\rm out}$ (see the sixth of six graphics in Figure \ref{fig:cartoon_ring}).
Thus, the no-ring area is given by
\begin{equation}\label{eq:S_F}
    S_{\rm no-ring,F}=\pi[ r^2 -(a_{\rm out}b_{\rm out}-a_{\rm in}b_{\rm in}) ].
\end{equation}
The derivative is simply given by
\begin{equation}\label{eq:dS_F/dr}
 \frac{dS_{\rm no-ring,F}}{dr}=2\pi r.
\end{equation}

\subsection{Summary of calculation procedures}\label{sec:method_summary}
In total we have six formulas for the derivative of the no-ring areas.
One can compute the transmission spectra of ringed planets with Equation \eqref{eq:spectrum_ring} by choosing an appropriate formula of $dS_{\rm no-ring}/dr$ at each radial distance $r$ along with Equation \eqref{eq:relation}.
For a ring with $a_{\rm in}>b_{\rm out}$, we never meet the range D and thus only need the ranges A, B, C, E, and F.
Therefore, the required formulas are Equations \eqref{eq:dS_A/dr}, \eqref{eq:dS_B/dr}, \eqref{eq:dS_C/dr}, \eqref{eq:dS_E/dr}, and \eqref{eq:dS_F/dr}.
For the ring with $a_{\rm in}<b_{\rm out}$, the range D applies instead of the range C.
Thus, the required formulas are Equations \eqref{eq:dS_A/dr}, \eqref{eq:dS_B/dr}, \eqref{eq:dS_D/dr}, \eqref{eq:dS_E/dr}, and \eqref{eq:dS_F/dr}.
In a specific case of the face-on ring with $a_{\rm in}=b_{\rm in}$ and $a_{\rm out}=b_{\rm out}$, we only meet the ranges A, D, and F.
Then, our model is reduced to the face-on ring model of \citet{Ohno&Tanaka21}.

\subsection{Postprocessing method to include the ring effects}\label{sec:approx}
It would be convenient if one can take the ring effect into account via postprocessing a precomputed ring-free spectrum.
Here, we propose such a simple postprocessing method.
Our idea is the reversal of the argument in the previous section: we no longer need to care about the details of the ring at the radial distance where the atmosphere is opaque.
We can define a threshold radial distance below which the atmosphere is completely opaque, which is essentially the same as the planetary effective radius $R_{\rm eff}$ computed by standard spectrum models, given by
\begin{equation}\label{eq:spectrum_R_eff}
    \pi R_{\rm eff}^2=\pi R_{\rm 0}^2+\int_{\rm R_{\rm 0}}^{R_{\rm s}} 2\pi r[1-\exp(-\tau)]dr.
\end{equation}
The effective radius is the radius of a rigid sphere whose transit depth is the same as the planet with an atmosphere.
The effective radius corresponds to the distance where the slant optical depth becomes $\approx 0.56$ \citep{Lecavelier+08,deWit&Seager13,Heng&Kitzmann17}.

Approximating the planet as an opaque disk with a radius of $R_{\rm eff}$, we can evaluate the observable radius of a ringed planet by adding the transmittance of the ring outside the planetary disk.
This procedure can be written by
\begin{equation}\label{eq:spectrum_post}
    \pi R_{\rm obs}^2=\pi R_{\rm eff}^2 + S_{\rm ring,out}(R_{\rm eff}) [1-\exp{(-\tau_{\rm ring,LOS})}],
\end{equation}
where $S_{\rm ring,out}$ is the ring's area outside the planetary disk, given by
\begin{equation}
    S_{\rm ring,out}(R_{\rm eff})=[\pi (a_{\rm out}b_{\rm out}-a_{\rm in}b_{\rm in})-S_{\rm ring}(R_{\rm eff})].
\end{equation}
The $S_{\rm ring}$ can be obtained from $S_{\rm no-ring}$ in previous sections (Equations \ref{eq:S_A}, \ref{eq:S_B}, \ref{eq:S_C}, \ref{eq:S_D},\ref{eq:S_E}) using a relation of $S_{\rm ring}(r)+S_{\rm no-ring}(r)=\pi r^2$.
Since the ring-free effective radius varies with wavelength, one can first compute $R_{\rm eff}$ on a given wavelength grid and then use Equation \eqref{eq:spectrum_post} at each wavelength.
The advantage of this method is that one can easily include the ring effects into the ring-free spectra computed by existing models \citep[e.g.,][]{Waldmann+15,Zhang+19,Molliere+19,Min+20,Cubillos&Blecic21} as well as precomputed grids \citep{Goyal+19} with negligible computational costs.
\revise{In Section \ref{sec:result_spectrum}, we will show that the postprocessing method can reproduce the spectra computed by the methods outlined in Section \ref{sec:method_summary}.}

\section{Limits on Ring Physical Properties}\label{sec:ring_property}
It has been often assumed that the ring is completely opaque in the analysis of photometric light curves \citep[e.g.,][]{Akinsanmi+20}.
However, the assumption of a completely opaque ring might require an unphysically massive ring in certain cases.
Furthermore, most of the previous studies assumed a gray ring opacity, while the actual ring opacity depends on the size distribution of ring particles.
In the subsequent sections, we discuss the possible physical properties of exoplanetary rings.

\subsection{Ring Surface Density}\label{sec:ring_surf}
Although the actual ring surface density is unknown for exoplanets, we can estimate the upper and lower limits of the surface density for optically thick rings.
If the ring surface density is too low, the ring would not be sustainable because the ring particles continuously fall to the planet owing to the Poynting-Robertson drag \citep{Gaudi+03,Schlichting&Chang11}.
The orbit decay timescale due to the Poynting-Robertson drag for the optically thick ring is given by \citep{Schlichting&Chang11}
\begin{equation}\label{eq:t_PR}
    \tau_{\rm PR}\sim \frac{\pi \Sigma c^2}{\sin{\theta}(5+\cos^2{\theta})}\frac{4\pi d^2}{L_{\rm *}},
\end{equation}
where $\Sigma$ is the mass surface density of the ring, $c$ is the light speed, $d$ is the planet's orbital distance, and $L_{\rm *}$ is the stellar luminosity.
The ring is sustainable if the timescale is longer than the ring's age $\tau_{\rm age}$.
Setting $\tau_{\rm PR}=\tau_{\rm age}$, we estimate the lower limit of ring surface density as
\begin{eqnarray}\label{eq:Sigma_min}
    \Sigma_{\rm min}&\sim&\frac{4\sigma T_{\rm eq}^4 \tau_{\rm age}}{\pi c^2}\sin{\theta}(5+\cos^2{\theta})\\
    \nonumber
    &\sim& 80~{\rm g~{cm}^{-2}} \left( \frac{T_{\rm eq}}{300~{\rm K}}\right)^{4}\left( \frac{\tau_{\rm age}}{1~{\rm Gyr}} \right),
\end{eqnarray}
where $\sigma$ is the Stefan-Boltzmann constant, and we have used the relation of $L_{\rm *}/4\pi d^2=4\sigma T_{\rm eq}^4$ with equilibrium temperature for zero Bond albedo $T_{\rm eq}$.
We have assumed $\theta=45$ deg for the second equality in Equation \eqref{eq:Sigma_min}. 
\revise{Thus, the second line of Equation \eqref{eq:Sigma_min} contains an order-of-unity uncertainty originating from the uncertain ring obliquity.}
It is interesting to note that the minimum surface density is mostly controlled by the planet's equilibrium temperature.

The ring with a surface density higher than Equation \eqref{eq:Sigma_min} can be sustainable.
However, a too massive ring is not sustainable because the self-gravity wakes cause efficient angular momentum transport, leading to a rapid spread out of ring particles to the planet and beyond the Roche radius \citep[e.g.,][]{Salmon+10}.
The ring's spreading timescale is given by \citep{Crida&Charnoz12}
\begin{equation}\label{eq:tau_nu1}
    \tau_{\rm \nu}=\frac{R_{\rm R}^2}{\nu(R_{\rm R})},
\end{equation}
where $R_{\rm R}$ is the Roche radius given by
\begin{equation}\label{eq:R_roche}
    R_{\rm R}=2.456 R_{\rm p} \left( \frac{\rho_{\rm p}}{\rho_{\rm r}}\right)^{1/3},
\end{equation}
$R_{\rm p}$ is the planetary radius, \revise{$\rho_{\rm p}$ is the planetary bulk density}, $\rho_{\rm r}$ is the internal density of ring particles, and $\nu$ is the ring's effective viscosity given by \citep{Daisaka+10}
\begin{equation}\label{eq:nu_ring}
    \nu \approx 26r_{\rm H}^{*5}\frac{G^2 \Sigma^2}{\Omega^3},
\end{equation}
where $G$ is the gravitational constant, $\Omega=\sqrt{GM_{\rm p}/r^3}$ is the Kepler angular velocity, and $M_{\rm p}$ is the planetary mass.
$r_{\rm H}^{*}$ is the ring particle's Hill radius normalized by the particle's diameter, given by \citep{Daisaka+10}
\begin{equation}
    r_{\rm H}^{*}=\frac{r}{2a}\left( \frac{2m_{\rm p}}{3M_{\rm p}}\right)^{1/3},
\end{equation}
\revise{where $a$ is the ring particle radius, and $m_{\rm p}=4\pi a^3\rho_{\rm r}/3$ is the mass of ring particles.}
The normalized Hill radius is simplified to $r_{\rm H}^{*}\approx 1.07275$ at the Roche radius \citep[see][]{Crida&Charnoz12}.
Combining Equation \eqref{eq:tau_nu1}, \eqref{eq:R_roche}, and \eqref{eq:nu_ring}, as shown by \citet{Crida&Charnoz12}, the spreading timescale can be written by the Kepler orbital period at the Roche radius and a ring-to-planet mass ratio, as
\begin{equation}\label{eq:tau_nu2}
    \tau_{\rm \nu}=\frac{2\pi}{\Omega(R_{\rm R})}\left( \frac{\pi}{52r_{\rm H}^{*5}}\right)\left( \frac{\pi R_{\rm R}^2\Sigma}{M_{\rm p}}\right)^{-2}.
\end{equation}
The viscous spreading becomes slow as the ring surface density decreases.
Thus, the viscous spreading would reduce the ring surface density until the spreading timescale becomes longer than the ring's age.
Setting $\tau_{\rm \nu}=\tau_{\rm age}$, we estimate the upper limit of ring surface density as
\begin{eqnarray}\label{eq:sigma_max}
    \Sigma_{\rm max}&=&\sqrt{\frac{\pi^2}{26 r_{\rm H}^{*5}\Omega(R_{\rm R})\tau_{\rm age} }}\frac{M_{\rm p}}{\pi R_{\rm R}^2}\\
    \nonumber
    &\sim& 130~{\rm g~{cm}^{-2}}~\left( \frac{M_{\rm p}}{10M_{\rm \oplus}}\right)^{1/3}\left( \frac{\rho_{\rm r}}{1~{\rm g~{cm}^{-3}}}\right)^{5/12}\left( \frac{\tau_{\rm age}}{1~{\rm Gyr}} \right)^{-1/2}.
\end{eqnarray}
We anticipate that the actual ring surface density resides between $\Sigma_{\rm min}$ and $\Sigma_{\rm max}$.

\begin{figure}[t]
\centering
\includegraphics[clip, width=\hsize]{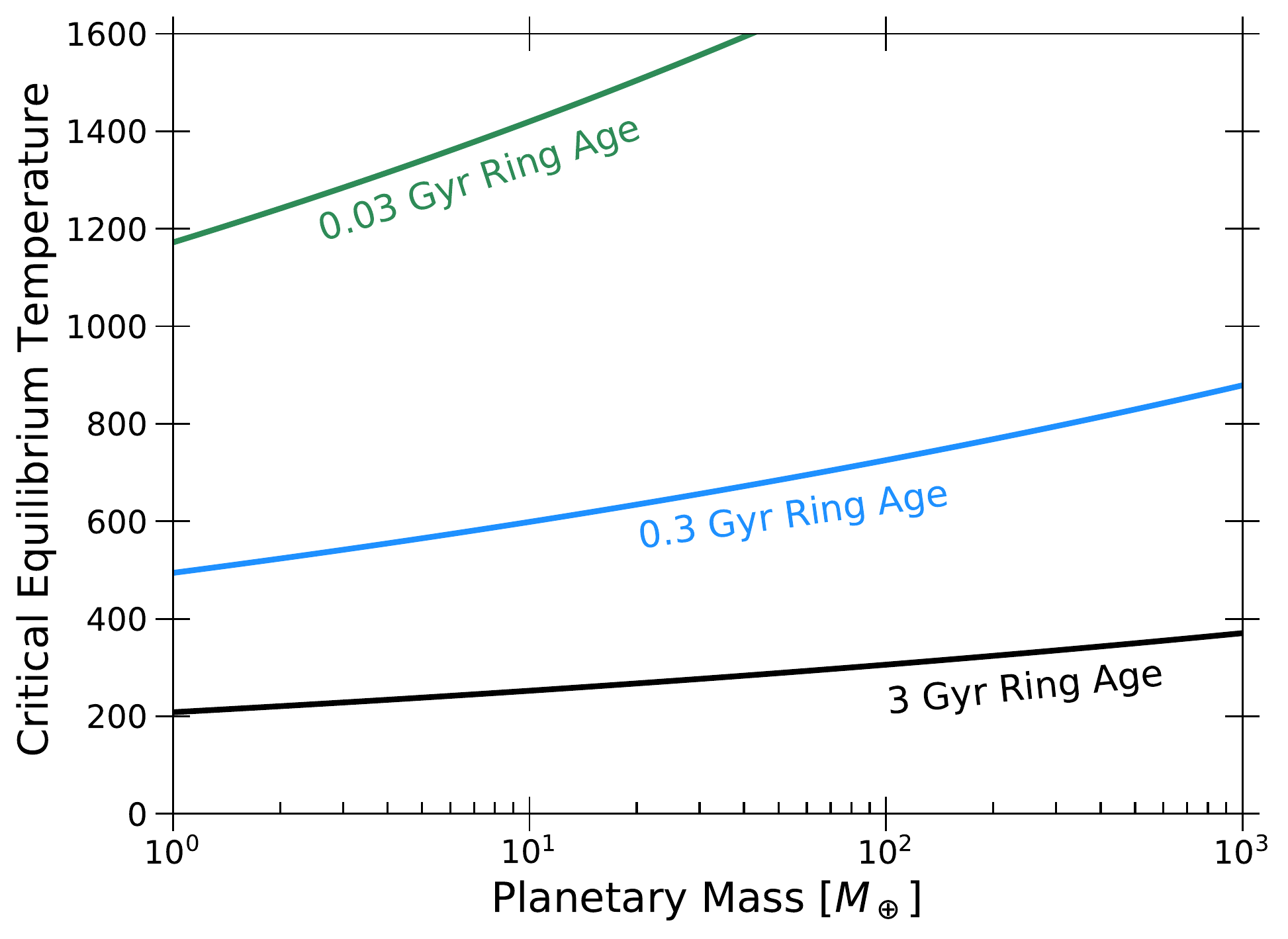}
\caption{Critical equilibrium temperature above which the optically thick ring may not be sustainable (see Section \ref{sec:ring_surf}). We assume the ring particle density of $\rho_{\rm r}=3~{\rm g~{cm}^{-3}}$.}
\label{fig:ring_limit}
\end{figure}

One can realize that there is a parameter space in which $\Sigma_{\rm min}>\Sigma_{\rm max}$.
In this regime, the ring continues spreading out until the ring becomes unsustainable owing to the Poynting-Robertson drag.
Thus, the optically thick ring may not be sustainable in that regime.
Solving $\Sigma_{\rm min}=\Sigma_{\rm max}$ with respect to $T_{\rm eq}$, we obtain a critical planetary equilibrium temperature above which the thick ring is unsustainable, given by
\begin{equation}\label{eq:T_cri}
    T_{\rm cri}\sim 340~{\rm K}~\left( \frac{M_{\rm p}}{10M_{\rm \oplus}}\right)^{1/12}\left( \frac{\rho_{\rm r}}{1~{\rm g~{cm}^{-3}}}\right)^{5/48}\left( \frac{\tau_{\rm age}}{1~{\rm Gyr}} \right)^{-3/8}.
\end{equation}
The essence of this analysis is the same as \citet{Goldreich&Tremaine79} who evaluated the maximum lifetime of the narrow ring of Uranus.
We plot Equation \eqref{eq:T_cri} in Figure \ref{fig:ring_limit}.
If the ring's age is $\sim3$ Gyr, which is comparable to the typical system age of {\it Kepler} planets \citep[][]{Berger+20}, the optically thick ring can be stable only when the equilibrium temperature is lower than $\sim300~{\rm K}$.
This might imply the intrinsic deficit of thick circumplanetary rings in the {\it Kepler} samples, which potentially explains the nondetection of short-period ringed planets in the {\it Kepler} sample \citep{Heising+15,Aizawa+18}.
However, we note a caveat that the ring's age is not necessary the same as the system age.
In fact, observations of the {\it Cassini} spacecraft suggest that the Saturnian ring may be much younger than Saturn itself, with an age of $0.01$--$0.1~{\rm Gyr}$ \citep{Zhang+17,Iess+19,Ida19}.
Thus, we rather suggest that the ring may be formed relatively recently if it is detected around close-in hot exoplanets.

\subsection{Ring Particle Size Distribution}\label{sec:ring_size}
Previous studies usually assumed gray ring opacity \citep[e.g.,][]{Aizawa+17}.
Gray ring opacity is a reasonable assumption since only $\gg1~{\rm cm}$ ring particles can survive over $0.1~{\rm Gyr}$ for known exoplanets owing to Poynting-Robertson drag \revise{when the ring is optically thin} \citep{Schlichting&Chang11}.
Saturnian rings consists of particles larger than $0.1~{\rm cm}$ \citep{Cuzzi+09}, and Uranian main rings mainly consist of particles larger than $\sim10~{\rm cm}$ \citep{Miner+07,DePater+13}.
Such large particles yield a gray opacity in the wavelength relevant to exoplanet observations.
However, as noted by \citet{Schlichting&Chang11}, tiny ring particles might survive if the ring is optically thick.
Moreover, the estimation of \citet{Schlichting&Chang11} for optically thin rings did not take into account the replenishment of tiny ring particles produced by collisional fragmentation of larger particles and the possible presence of moons, which may continuously supply fresh tiny ring particles.
In the solar system, the tenuous rings of Jupiter mainly consist of $\sim0.1$--$100~{\rm {\mu}m}$ particles \citep{Showalter+87}, Neptune also has the rings consisting of micron-sized particles \citep{DePater+18}, and Uranus has tiny dust rings between the planet's main rings \citep{DePater+13}.

Given the large uncertainty of exoplanetary rings, it is worth examining the impact of tiny ring particles on the observable transmission spectra.
In general, the ring optical depth can be calculated by
\begin{equation}
    \tau_{\rm ring}=\int_{a_{\rm min}}^{a_{\rm max}}Q_{\rm ext}\pi a^2(a)\mathcal{N}(a)da,
\end{equation}
where $Q_{\rm ext}$ is the extinction coefficient, $a$ is the ring particle radius, $\mathcal{N}(a)$ is the column-integrated size distribution of ring particles, and $a_{\rm min}$ and $a_{\rm max}$ are the smallest and largest radii of ring particles.
The column number density is associated with the ring surface density as
\begin{equation}
    \Sigma = \int_{a_{\rm min}}^{a_{\rm max}}m_{\rm p}\mathcal{N}(a)da,
\end{equation}
Here, we conventionally assume a power-law size distribution \citep[e.g.,][]{Cuzzi+09} given by
\begin{equation}
    \mathcal{N}(a)=Ca^{-\gamma},
\end{equation}
where $C$ is the constant determined from the surface density, as
\begin{equation}
    C=\frac{3(4-\gamma)\Sigma}{4\pi \rho_{\rm r}(a_{\rm max}^{4-\gamma}-a_{\rm min}^{4-\gamma})}.
\end{equation}
Theoretical studies showed that the universal power-law index of $\gamma=2.75$--$3.5$ exists for steady-state size distributions controlled by aggregation-fragmentation equilibrium \citep[e.g.,][]{Tanaka+96,Birnstiel+11,Brilliantov+15}. 
This is also in agreement with the size distribution of the Saturnian ring \citep[$\gamma\approx3$,][]{Cuzzi+09}, Jovian ring \citep[$\gamma=2.5\pm0.5$,][]{Showalter+87}, and Uranunian ring \citep[$\gamma=2.5\pm0.5$,][]{Ockert+87}.
In Section \ref{sec:nongray}, we will test the impact of non-gray ring opacity assuming $\gamma=3$, similar to Saturnian ring particles.
\revise{We leave the examination of more complex size distributions, such as the broken power-law suggested for the Saturn's F ring \citep{Hedman+11}, to future studies.}

We find that one can understand the universal power-law index of $\gamma=2.75$--$3.5$ from a dimensional argument.
Let us consider the particle mass distribution of $n(m)~[{\rm L^{-3}~M^{-1}}]$, where $n(m)dm$ is the number density of particles with masses between $m$ and $m+dm$.
The particle mass density within the mass grid between $m$ and $m+dm$ is thus $mn(m)dm$.
In steady state, the particle mass flux in a mass-space, $\dot{m}mn(m)~[{\rm M~L^{-3}~T^{-1}}]$, should be constant because the collision conserves the particle mass.
Since a collision Kernel $K=Av$, where $A$ and $v$ are the collision cross section and relative velocity, has the dimension of $[{\rm L^{3}~T^{-1}}]$, the form of the mass flux through a mass space can be speculated as
\begin{equation}\label{eq:dimension1}
    \dot{m}mn(m) \sim K m^3n(m)^2,
\end{equation}
Assuming the power-law mass distribution of $n(m)\propto m^{-\psi}$ and collision velocity $v\propto m^{-\xi}$, one can find the mass dependence of the flux as
\begin{equation}
    \dot{m}mn(m) \propto m^{11/3-\xi-2\psi}.
\end{equation}
Since the steady state demands a constant mass flux thorough mass space, $\psi$ should satisfy
\begin{equation}
   \psi =\frac{11-3\xi}{6}.
\end{equation}
Using the relation of $n(a)=n(m)dm/da$, one can obtain the power-law index of size distribution as
\begin{equation}
    n(a)\propto m^{2/3-\psi} \propto a^{-(7-3\xi)/2}.
\end{equation}
Thus,
\begin{equation}
\gamma = \frac{7-3\xi}{2}.
\end{equation}
When the collision velocity is independent of the particle mass, i.e., $\xi=0$, we recover the index of $\gamma=3.5$.
That said, if we assume the law of energy equipartition, i.e., $(1/2)m\langle{v}\rangle^2={\rm const}$ and thus $\xi=1/2$, we obtain $\gamma=2.75$ as derived in \citep{Brilliantov+15}.
For more elaborated discussions, we refer readers to \citet{Tanaka+96}, \citet{Birnstiel+11}, and \citet{Brilliantov+15}.

\section{Results}\label{sec:results}
In this section, we show how rings affect the transmission spectra.
For the sake of simplicity, we assume $\varphi=0$ (see Figure \ref{fig:cartoon_ring}) that allows the face-on ring geometry. 
Then, the viewing geometry can be characterized by the sky-plane angle $\phi$, which is associated with the ring's obliquity as $\phi=\pi/2-\theta$.

\subsection{Atmospheric contribution fraction}
\begin{figure}[t]
\centering
\includegraphics[clip, width=\hsize]{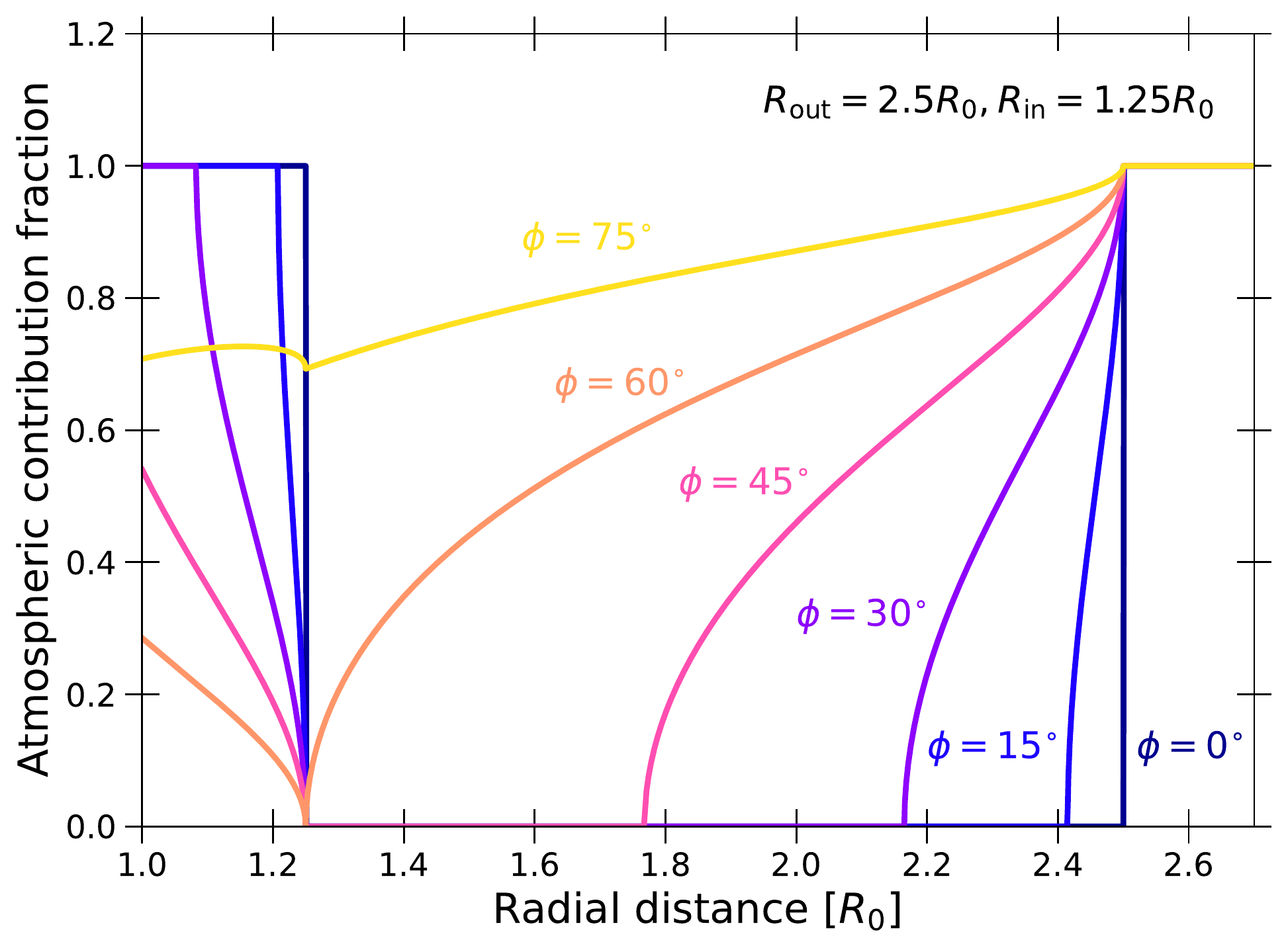}
\caption{Atmospheric contribution fraction, defined as $(dS_{\rm no-ring}/dr)/2{\pi}r$, as a function of radial distance from the planet center. Different colored lines show the contribution fraction for different sky-plane angle $\phi$. 
We assume $R_{\rm in}=1.25R_{\rm 0}$ and $R_{\rm out}=2.5R_{\rm 0}$.
}
\label{fig:atm_contribution}
\end{figure}
\begin{figure*}[t]
\centering
\includegraphics[clip, width=\hsize]{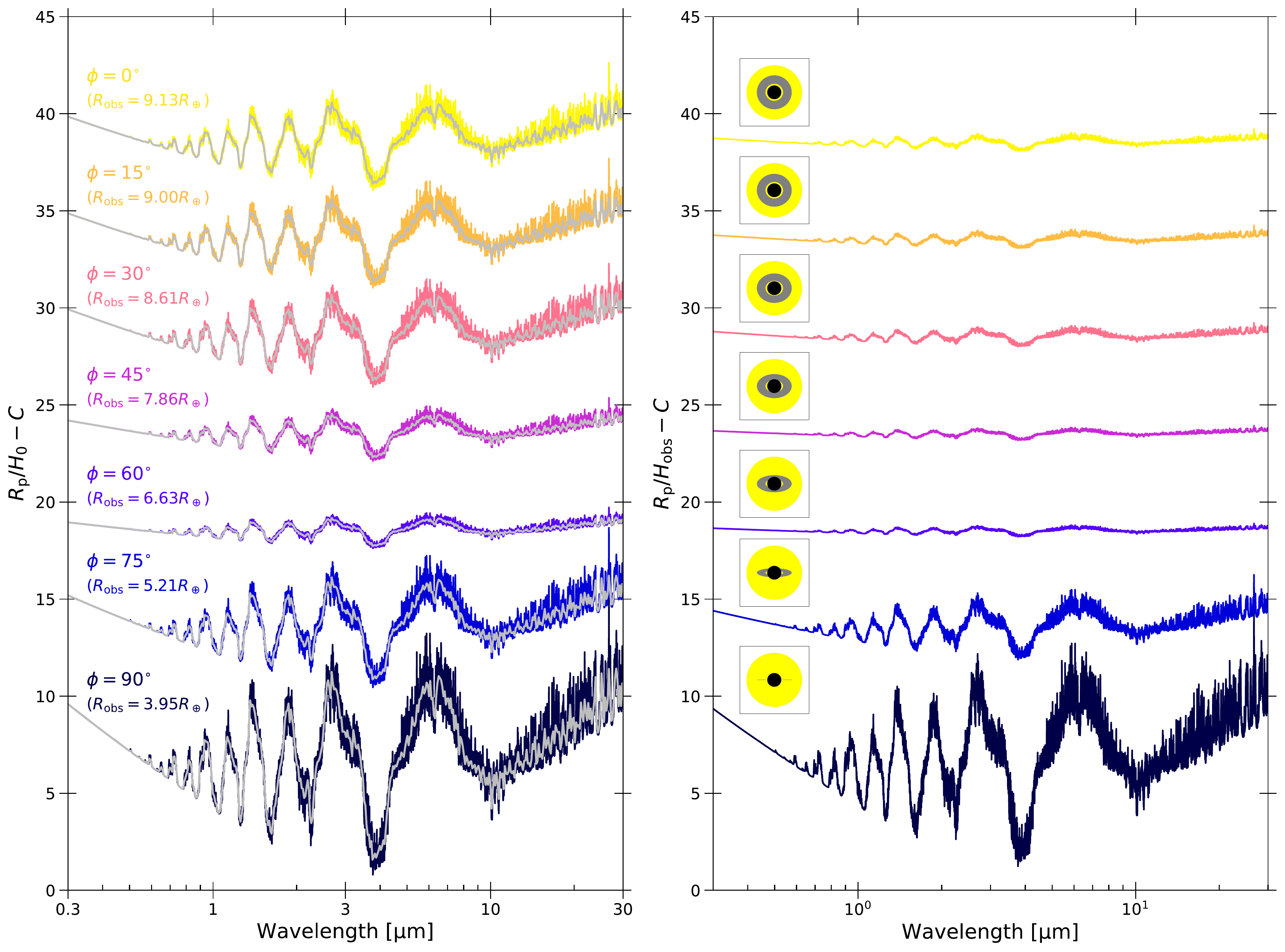}
\caption{Transmission spectra of ringed planets for various sky-plane angles. Different colored lines show the transit radii for different ring sky-plane angle $\phi$.
The \revise{silver} thin lines in the left panel show the spectra computed by the postprocessing method introduced in Section \ref{sec:approx}, which are binned down to a lower resolution for clarity, and show excellent agreement with the detailed transmission spectrum calculation (colors).
We assume $M_{\rm p}=10M_{\rm \oplus}$ and $R_{\rm 0}=3.8R_{\rm \oplus}$.
The transit radii \revise{at the wavelength of $\lambda=0.6~{\rm \mu m}$, the center of the Kepler bandpass,} is also denoted.
The transit radii are normalized by the `true' pressure scale height $H_{\rm 0}$ computed at $r=R_{\rm 0}$ in the left panel with arbitrarily offsets, while the right panel shows the radii normalized by the `observed' scale height $H_{\rm obs}$ computed at $r=R_{\rm obs}$.
\revise{The right panel also involves a cartoon illustrating the appearance of ringed planet, where black circle represents the planetary disk with a radius of $R_{\rm 0}$, and gray part denotes the ring.}
}
\label{fig:spectrum1}
\end{figure*}

The derivative area tells us how much the atmospheric feature contributes to the spectrum at each radial distance.
Figure \ref{fig:atm_contribution} shows the atmospheric contribution fraction, which we define by $(dS_{\rm no-ring}/dr)/2{\pi}r$, at each radial distance for various sky-plane angle $\phi$ and ring inner edges.
A contribution fraction of unity means that the whole annulus of the atmosphere is observable, while a value of zero means that the whole annulus is overlapped by the ring.
A face-on ring ($\phi=0^{\circ}$) always has an atmospheric contribution of unity at $r<R_{\rm in}$ and $r>R_{\rm out}$ along with a zero contribution at $R_{\rm in}<r<R_{\rm out}$, as assumed in \citet{Ohno&Tanaka21}.
The atmospheric contribution at $r<R_{\rm in}$ gradually decreases as the sky-plane angle $\phi$ increases.
This is because a higher $\phi$ leads the projected ring's inner edge to be placed closer to the planet.
The atmospheric contribution at $r=R_{\rm 0}$ always becomes lower than unity at $\phi>\cos^{-1}{(R_{\rm 0}/R_{\rm in})}$, corresponding to $\sim36^{\circ}$ here, since an atmospheric annulus at $r>R_{\rm 0}$ always involves the ring-overlapped area. 
Conversely, the atmospheric contribution at $R_{\rm in}<r<R_{\rm out}$ increases with increasing $\phi$. 
This is because a higher $\phi$ also leads the projected outer edge to be closer to the planet, which in turn creates no-ring areas in the atmospheric annulus at $r>R_{\rm out}\cos{\phi}$. 

The atmospheric contribution at around the planetary radius $r\sim R_{\rm 0}$, from which the atmospheric features mainly originate, becomes a minimum at $\phi\sim60^{\circ}$.
This corresponds to a threshold angle of $\phi_{\rm cri}=\cos^{-1}{(R_{\rm in}/R_{\rm out})}$ below which the range C never meets (see Figure \ref{fig:cartoon_ring}). 
In particular, the atmospheric contribution at $r\sim R_{\rm 0}$ always decreases with increasing the sky-plane angle at $\cos^{-1}{(R_{\rm 0}/R_{\rm in})}<\phi<\phi_{\rm cri}$ because the ring-free area in an annulus around the planetary radius monotonically decreases with increasing $\phi$.
Conversely, as seen in the $\phi=75^{\circ}$ case, the atmospheric contribution at $r\sim R_{\rm 0}$ becomes relatively high at very high $\phi$ corresponding to a nearly edge-on ring.
This is because the area of the projected ring itself decreases with increasing $\phi$.
In particular, the ring-overlapped area in the annulus around $r\sim R_{\rm 0}$ always decreases with increasing the sky-plane angle at $\phi>\cos^{-1}{(R_{\rm 0}/R_{\rm out})}$ ($\sim66^{\circ}$ here), as the width of the projected ring is smaller than the planetary reference radius.
Thus, a ring with an intermediate inclination, especially at $\phi_{\rm cri}\le \phi\le\cos^{-1}{(R_{\rm 0}/R_{\rm out})}$, can most efficiently mask the atmosphere.

\subsection{Effects of Ring on Transmission Spectra}\label{sec:result_spectrum}

In this section, we investigate how the ring properties affect the transmission spectra.
We compute the transmission spectrum with the model of \citet{Ohno&Tanaka21} only including the absorption of H$_2$O and the Rayleigh scattering by H$_2$ and He.
For the sake of simplicity, we assume an isothermal atmosphere with a temperature of $T=300~{\rm K}$ and with a mean molecular weight of $2.35~{\rm amu}$.
We consider a planet with a mass of $10M_{\rm \oplus}$ and reference radius of $3.8R_{\rm \oplus}$, similar to the $10M_{\rm \oplus}$ mass planet with $10\%$ atmospheric mass fraction at the system age of $1~{\rm Gyr}$ \citep{Lopez&Fortney14}.
We set the atmospheric pressure at the planetary reference radius $R_{\rm 0}$ to be $30~{\rm bar}$.
In addition to the method outlined in Sections \ref{sec:method}--\ref{sec:method_summary}, we also compute the spectra following the postprocessing method introduced in \ref{sec:approx} to test its validity.
In this section, we assume a completely opaque ring.

The transmission spectrum varies considerably with the viewing geometry of the circumplanetary ring.
The left panel of Figure \ref{fig:spectrum1} shows the transit radii of ringed planets normalized by the `true' atmospheric scale height, the scale height computed at $r=R_{\rm 0}$, for various sky-plane angles.
The ring has the outer and inner edge radii of $R_{\rm out}=2.5R_{\rm 0}$ and $R_{\rm in}=1.25R_{\rm 0}$, respectively, in Figure \ref{fig:spectrum1}.
In general, the ring acts to suppress the spectral features.
The planet with an edge-on ring of $\phi=90^{\circ}$ shows the prominent H$_2$O features ranging across several atmospheric scale heights, as the edge-on ring is essentially the ring-free case owing to the neglected ring thickness.
As the ring gets inclined to the orbital plane, the spectral features get more obscured because a part of the planetary atmosphere is completely overlapped with the ring (see also cartoon in the right panel).
The spectral feature amplitudes are minimized at around $\phi={60}^{\circ}$ because the atmospheric contribution fraction near the reference radius $R_{\rm 0}$ is the lowest in this sky-plane angle (see Figure \ref{fig:atm_contribution}). 
We note that the atmospheric features are mainly coming from the inner hole of the projected ring, as the scale height is much smaller than the length scale of the planet and the ring.
As the ring has a more face-on geometry, the spectral features become stronger because the projected ring's inner edge is apart from the reference radius $R_{\rm 0}$ (see cartoon in the right panel).
In the limit of a face-on ring ($\phi=0^{\circ}$), the spectrum shows moderate H$_2$O features, although they are still weaker than the ring-free spectra.

\revise{
One might wonder why the spectral features for the face-on ring are still weaker than those for the edge-on ring, i.e., a ring-free spectrum.
This stems from the fact that the observed radius is not linear in the occultation area.
The amplitude of spectral features can be quantified by the wavelength derivative of the observed radius.
Differentiating Equation \eqref{eq:spectrum_post} with respect to wavelength yields
\begin{equation}\label{eq:dRdlam}
\frac{dR_{\rm obs}}{d\lambda}=\frac{\left( \frac{d R_{\rm eff}}{d\lambda}+\frac{S_{\rm ring,out}e^{-\tau_{\rm ring,LOS}}}{2R_{\rm eff}}\frac{d \tau_{\rm ring,LOS}}{d\lambda} \right)}{\left[ 1+\frac{S_{\rm ring,out}(1-e^{-\tau_{\rm ring,LOS}})}{\pi R_{\rm eff}^2} \right]^{1/2}},
\end{equation}
where we recall that $R_{\rm eff}$ is the effective radius of the ring-free planet.
If we assume $\tau_{\rm ring,LOS}\gg 1$ as in Figure \ref{fig:spectrum1}, the equation simplifies to
\begin{equation}
\frac{dR_{\rm obs}}{d\lambda}\approx\frac{d R_{\rm eff}}{d\lambda}\left( 1+\frac{S_{\rm ring,out}}{\pi R_{\rm eff}^2} \right)^{-1/2}.
\end{equation}
Thus, the spectral features in terms of the observed radius, is always weaker than those of ring-free planets, $dR_{\rm eff}/d\lambda$, by a factor of $(1+S_{\rm ring,out}/\pi R_{\rm eff}^2)^{1/2}$.
}

We here stress that observers do not know the `true' atmospheric scale height in practice, as the reference radius $R_{\rm 0}$ is much smaller than the observed radius $R_{\rm obs}$ for ringed planets.
This fact further changes the interpretation of the observed spectra.
The right panel of Figure \ref{fig:spectrum1} shows the same transit radii of the ringed planets, but normalized by the `observed' atmospheric scale height $H_{\rm obs}$ calculated from the wavelength-averaged transit radius. 
If we assume the ring extends from near the planetary true radius to the Roche radius of Equation \eqref{eq:R_roche}, crudely speaking, the observed scale height overestimates the true scale height by a factor of $\sim6\cos{\phi}$.
As a result, as seen in Figure \ref{fig:spectrum1}, the spectra are nearly flat at $\phi\la60^{\circ}$.
Thus, the combination of the masked atmosphere and the overestimated scale height results in fairly featureless spectra with amplitudes of only $\la H_{\rm obs}$, unless the ring is nearly edge-on.

In the left panel of Figure \ref{fig:spectrum1}, we also plot the spectra computed by the postprocessing method introduced in Section \ref{sec:approx}.
We find that the postprocessing method successfully reproduces the spectra computed by properly accounting for the ring-free and -overlapped areas of each atmospheric annulus.
Thus, the postprocessing method will be useful as an efficient way to interpret the spectra of anomalously low-density exoplanets.

\begin{figure}[t]
\centering
\includegraphics[clip, width=\hsize]{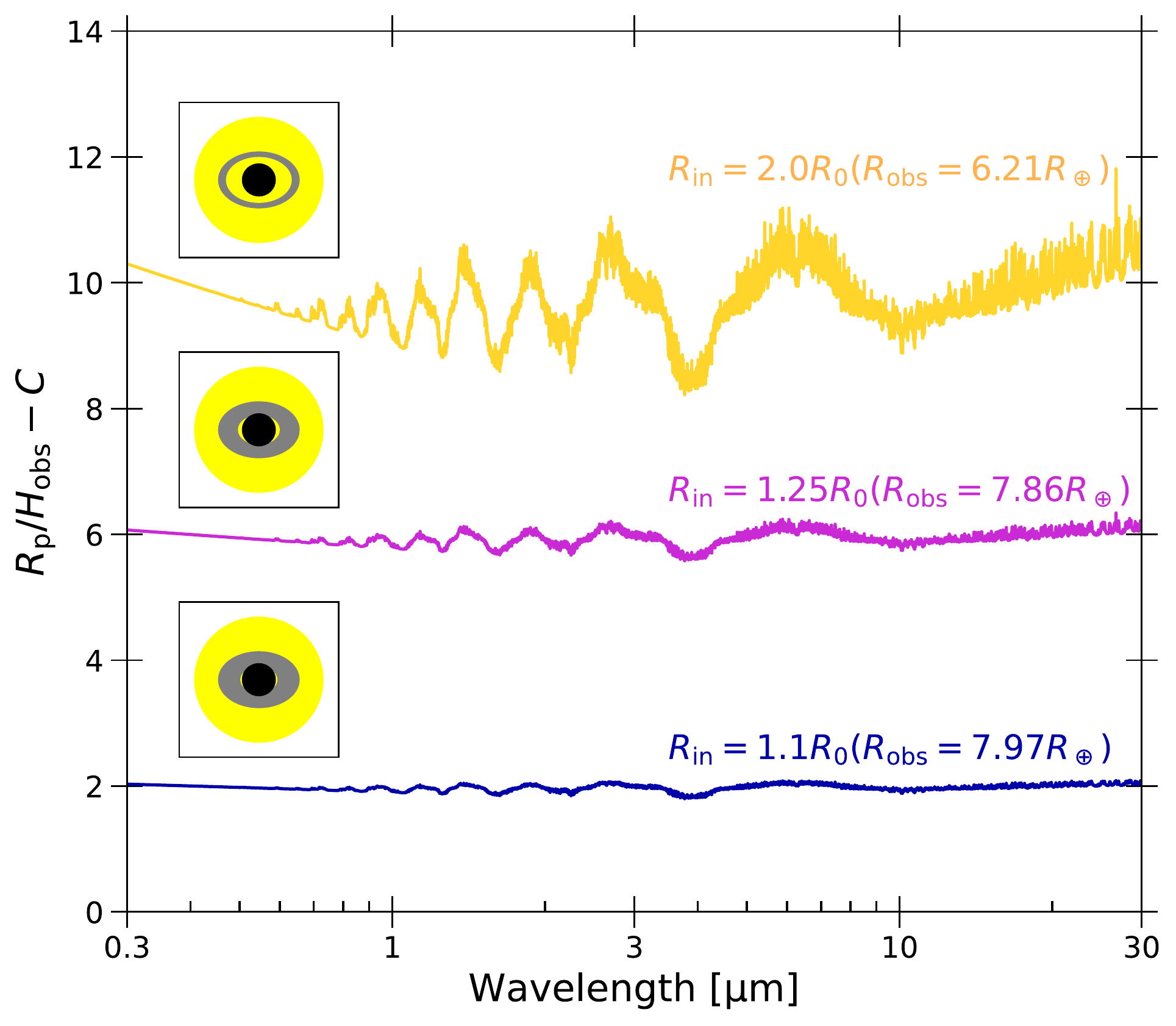}
\caption{Same as Figure \ref{fig:spectrum1}, but for different ring's inner edge radii. 
The transit radii are normalized by the observed scale height.
We fix the sky-plane angle to $\phi={45}^{\circ}$.
}
\label{fig:spectrum_Rin}
\end{figure}

The flatness of the transmission spectrum also depends on where the ring's inner edge is.
Figure \ref{fig:spectrum_Rin} shows the transmission spectra for the inner edge radii of $R_{\rm in}=1.1$, $1.25$, and $2.0R_{\rm 0}$.
The closer to the planet the ring's inner edge is, the flatter the spectrum is.
This is because the closer inner edge can more easily mask the ring's inner hole where the atmospheric features mainly come from.
When the ring's inner edge is far away from the planet, the ring barely obscures the atmospheric features as seen in the $R_{\rm in}=2R_{\rm 0}$ case.
Thus, a nearly flat observed spectrum potentially indicates the presence of a broad Saturn-like ring rather than a confined Uranus-like ring, if the ring indeed exists.

\subsection{Can We See the Spectral Feature of the Ring Itself?}\label{sec:nongray}
\begin{figure}[t]
\centering
\includegraphics[clip, width=\hsize]{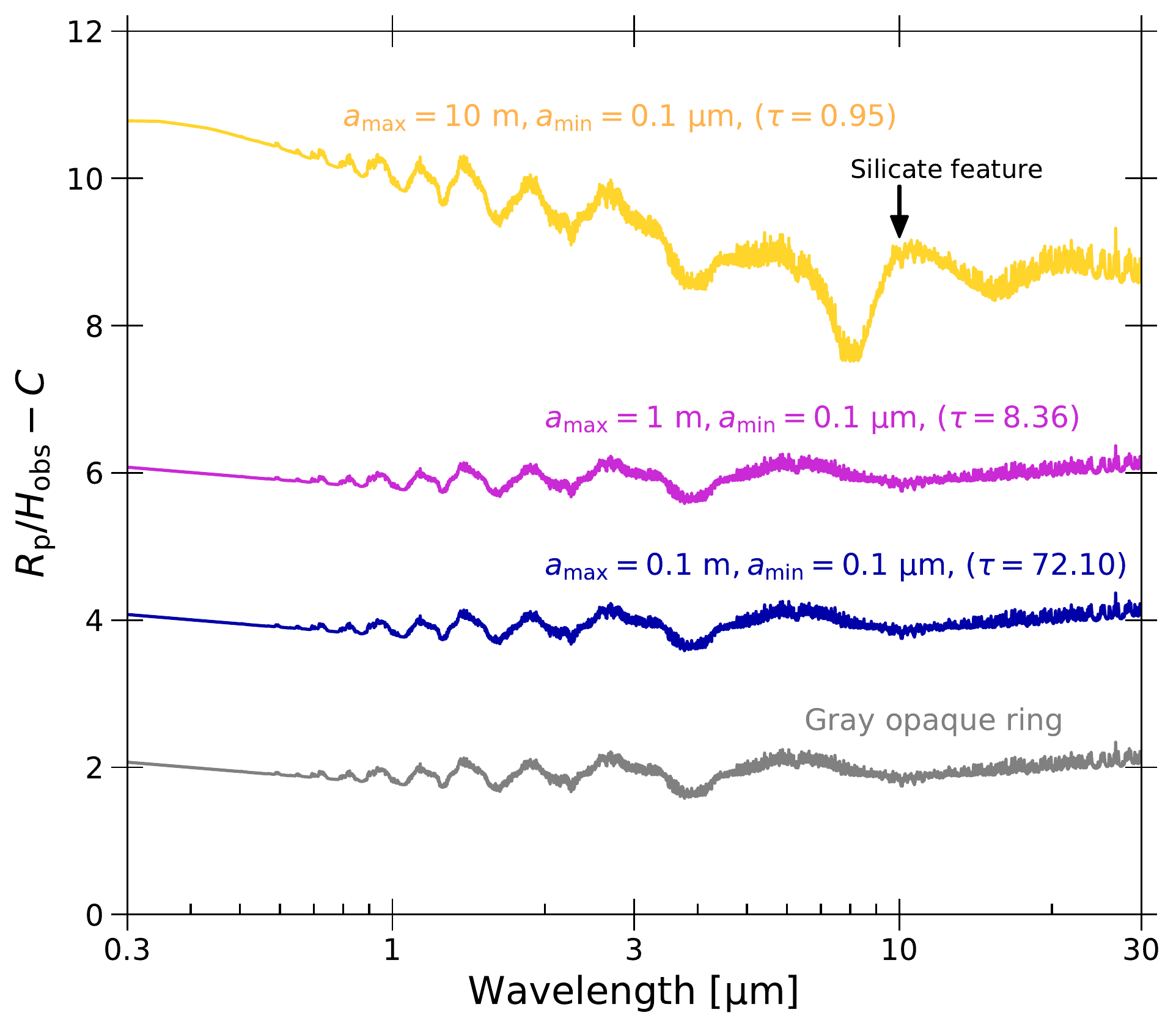}
\caption{Transmission spectra with a ring opacity computed from a particle size distribution. Different colored lines show the spectra for different largest sizes of ring particles. We assume $\phi=45^{\circ}$, $R_{\rm in}=1.25R_{\rm 0}$, $R_{\rm out}=2.5R_{\rm 0}$, $\Sigma=100~{\rm g~{cm}^{-2}}$, $\rho_{\rm r}=3~{\rm g~{cm}^{-3}}$, $\gamma=3$, and $a_{\rm min}=0.1~{\rm {\mu}m}$. For a comparison, we also plot the spectrum with gray opaque ring as a gray line. We also label the line-of-sight ring optical depth at $\lambda=1~{\rm {\mu}m}$.
}
\label{fig:spectrum_nogray}
\end{figure}
While we assumed a gray opaque ring thus far, in reality, the rings have non-gray opacity depending on the particle properties.
We test the impact of the non-gray ring opacity assuming Saturn like rings with $\gamma=3$ and $\Sigma=100~{\rm g~{cm}^{-2}}$.
We calculate the non-gray ring spectrum with the postprocessing method of Section \ref{sec:approx}: we first compute the ring optical depth at each wavelength with Mie theory \citep{Bohren&Huffman83} and then use Equation \eqref{eq:spectrum_post}.
The refractive index used is taken for astronomical silicate \citep{Draine03}, as the known possible ringed exoplanets are too warm to sustain icy rings \citep{Piro&Vissapragada19}.
Thus, we assume the ring particle internal density of $\rho_{\rm r}=3~{\rm g~{cm}^{-3}}$.
If the smallest particle size is $a_{\rm min}\sim0.1~{\rm cm}$ as in Saturnian rings, the ring opacity becomes almost gray at infrared wavelengths.
However, since the size distribution of exoplanetary rings is unknown, we instead choose $a_{\rm min}=0.1~{\rm {\mu}m}$ to maximize the potential impact of the non-gray ring opacity.
We vary the largest particle size $a_{\rm max}$ from $0.1$ to $10~{\rm m}$.
We note that the Saturnian ring particles have a largest size of $\sim3$--$20~{\rm m}$ \citep{Cuzzi+09}.

We find that non-gray ring opacity affects the transmission spectrum only when the ring has an optical depth around unity.
Figure \ref{fig:spectrum_nogray} shows the transmission spectra for various largest particle sizes $a_{\rm max}$.
The spectra with $a_{\rm max}=0.1$ and $1~{\rm m}$ are almost the same as the spectrum computed with a gray opaque ring, while the spectrum with $a_{\rm max}=10~{\rm m}$ shows a spectral slope and silicate features around $\lambda\sim10~{\rm {\mu}m}$.
This result might seem counter intuitive, as the fractional contribution of tiny particles to the ring opacity is lower for larger $a_{\rm max}$.
The reason originates from the fact that the ring's occultation area is limited by its physical size.
Once the ring's optical depth largely exceeds unity, the ring's occultation area is nearly the projected ring's area regardless of the optical depth.
In other words, we cannot realize the wavelength dependence of $\tau_{\rm ring}$ from the ring's occultation area when $\tau_{\rm ring}\gg1$.
This is why the spectra for $a_{\rm max}=0.1$ and $1~{\rm m}$, which have optical depth much higher than unity, are nearly the same as the gray model.
\revise{We can also understand this fact from Equation \eqref{eq:dRdlam}, where the term of wavelength derivative of the ring optical depth becomes negligible at $\tau_{\rm ring}\gg1$.}
On the other hand, the ring with $a_{\rm max}=10~{\rm m}$ has an optical depth about unity, which enables us to see the wavelength dependence of the ring's opacity in the variation of the ring's occultation area.
Thus, it would be possible to give tight constrains on the particle properties of the ring from the requirement of $\tau_{\rm ring}\sim1$ if one observes the ring's silicate features in the transmission spectrum.
The potential presence of the silicate feature will be testable by the \emph{JWST} MIRI \revise{and potentially help to distinguish the ring from other extra opacity sources, such as aerosols, as discussed in Section \ref{sec:discussion2}}.

\section{Discussion}\label{sec:discussion}
\subsection{Implications for the Young Exoplanet Observations}
We anticipate that younger planets have better chances to retain massive observable rings.
As shown in Figure \ref{fig:ring_limit}, even hot Jupiters can retain optically thick rings at the system age of $<100~{\rm Myr}$.
A number of such young exoplanets have already been discovered recently, such as AU Mic b \citep{Plavchan+20,Hirano+20}, V1298 Tau b \citep{David+19}, Ds Tuc Ab \citep{Newton+19}, TOI 1227 b \citep{Mann+21}, HIP 67522b \citep{Rizzuto+20}, and K2-33b \citep{Mann+16}.
Some of them appear to be large as compared to the {\it Kepler} planets \citep{Mann+16,Mann+16b,Obermeier+16}.
Although young exoplanets are intrinsically large owing to the postformation heat in the planetary interior \citep[e.g.,][]{Fortney+07,Lopez&Fortney14}, the potential presence of the ring may also be responsible for the large transit radii of young exoplanets.


\subsection{How to Distinguish Ringed Low-density Exoplanets from Other Possibilities}\label{sec:discussion2}
The presence of rings is not a unique solution for explaining the extremely low bulk density of exoplanets.
A large atmospheric mass fraction naturally causes inflated planetary radii \citep[e.g.,][]{Lopez&Fortney14}, which may originate from planet formation far away from a planet's current orbit \citep{Lee&Chiang16,Chachan+21}.
Strong interior heating can also cause the inflated radius \citep{Millholland19}.
Atmospheric aerosols can be responsible for both the anomalously large transit radius and featureless transmission spectrum \citep[e.g.,][]{Wang&Dai19,Gao&Zhang20,Ohno&Tanaka21}, similar to ring scenarios.

Several observational tests will help break the degeneracy between the different origins of extremely low-density exoplanets.
The current microphysical models predict that aerosols cause a strongly sloped featureless spectrum \citep[e.g.,][]{Gao&Zhang20}, which likely causes a noticeable difference from a nearly flat ring spectrum at a wider wavelength coverage \citep{Ohno&Tanaka21}.
\emph{JWST} would be able to see such a difference, along with seeing if the ring's silicate feature is present.
An intrinsically inflated exoplanet due to a high atmospheric mass fraction and/or interior heating will have a surface gravity that is much lower than that of a ringed planet.
Thus, an intrinsically inflated planet would yield prominent spectral features in the transmission spectrum if the planet has a haze-free atmosphere \citep[see e.g.,][]{Kawashima+19}, although this is inconsistent with the current observations.
Even if the inflated planet also has a cloudy/hazy atmosphere being consistent with the current observation, such a planet should undergo atmospheric escape that is much stronger than ringed planets.
Thus, a ringed exoplanet may be discovered as an extremely low density planet that exhibits unexpectedly weak atmospheric escape. 

Apart from the transmission spectrum, various studies have proposed the potential observable signatures of rings, such as the diffraction of star light by ring particles \citep{Barnes&Fortney04}, radial velocity anomaly \citep{Ohta+09}, and distortion of stellar lines \citep{deMooij+17}.
These observations can be carried out by current and future ground-based telescopes.
Complementary observational signatures would help to conclusively identify exoplanetary rings.

\subsection{Possible Model Extensions to Multiple Rings}
In this study, we have assumed a single ring for the sake of simplicity.
In reality, the planet can be surrounded by multiple rings, as in the rings of solar system giant planets.
It will be straightforward to extend our framework to the multiple ring system. 
If the planet is surrounded by $n$ rings, we can generalize Equation \eqref{eq:spectrum_ring} as
\begin{eqnarray}\label{eq:spectrum_ring_extend}
    D = \frac{1}{\pi R_{\rm s}^2}[&&\pi R_{\rm 0}^2+ \int_{\rm R_{\rm 0}}^{R_{\rm s}} [1-\exp{(-\tau)}]\frac{dS_{\rm no-ring}}{dr}dr\\
    \nonumber
    &&+ \sum_{\rm i}^{\rm n}\int_{\rm R_{\rm 0}}^{R_{\rm s}}[1-\exp{[-(\tau+\tau_{\rm ring,LOS,i})]}]\frac{dS_{\rm ring,i}}{dr}dr ],
\end{eqnarray}
where $\tau_{\rm ring,LOS,i}$ is the line-of-sight optical depth of the i-th ring, $dS_{\rm ring,i}/dr$ is the derivative of the i-th ring's area, and $dS_{\rm no-ring}/dr$ is calculated by
\begin{equation}
    \frac{dS_{\rm no-ring}}{dr}=2\pi r-\sum_{\rm i}^{\rm n}\frac{dS_{\rm ring,i}}{dr}.
\end{equation}
At each radial distance, one can use the method outlined in Section \ref{sec:method}--\ref{sec:method_summary} to calculate a {\it no i-th ring area} that can be converted to the derivative of i-th ring's area through the relation of $dS_{\rm ring,i}/dr=2\pi r-dS_{\rm no-ring,i}/dr$.
The postprocessing method can also be generalized as
\begin{equation}\label{eq:spectrum_R_eff_extend}
    \pi R_{\rm obs}^2=\pi R_{\rm eff}^2 + \sum_{\rm i}^{\rm n}S_{\rm ring,out,i}(R_{\rm eff}) [1-\exp{(-\tau_{\rm ring,LOS,i})}].
\end{equation}
Because each projected ring does not overlap each other, one can simply sum up the transmittance of each ring.
We note that the above argument is based on the assumption that all rings have the same spin axis.
If the rings have own spin axis which differs from each other, the computation can be more complicated because the projected rings can be overlapped with each other.
Such complication is beyond the scope of this study.

\subsection{Caveat for the Postprocessing Method}\label{sec:caveat_postprocess}
Our postprocessing method \revise{introduced in Section \ref{sec:approx}} implicitly assumes homogeneous planetary terminators so that the planet can be regarded as a circular disk.
This assumption may no longer hold for close-in exoplanets since previous studies suggest noticeable inhomogeneties in exoplanetary day-night terminators from theoretical \citep[e.g.,][]{Fortney+10,Burrows+10,Line&Parmentier16,Kempton+17,Powell+19,Helling+19,Steinrueck+21,Espinoza&Jordan21,Welbanks&Madhusudhan21} and observational point of view \citep[e.g.,][]{Macdonald&Madhusudhan17a,Pinhas+19}.
If the atmospheric profile is significantly different between the morning and evening terminators \revise{in terms of the pressure-temperature profile as well as cloud and haze properties}, the postprocessing method might cause errors. 
For example, the postprocessing method might underestimate the ring impacts on the extended hot evening terminator, as the transit radius of the evening terminator is reduced to a limb-averaged smaller radius in the postprocessing method, \revise{as illustrated in the top panel of Figure \ref{fig:limb_spectrum}}.
Our general method outlined in Section \ref{sec:method}--\ref{sec:method_summary} will still be applicable if \revise{each terminator is uniform in latitude}.
\revise{With homogeneous terminators}, one can apply $(dS_{\rm no-ring}/dr)/2$ and $(dS_{\rm ring}/dr)/2$ for each morning and evening terminator thanks to the point symmetry of the projected ring.
We note that thick circumplanetary rings will be more likely detected for cool, non-tidally locked exoplanets (Section \ref{sec:ring_surf}).
Thus, we expect that the postprocessing method \revise{of Section \ref{sec:approx}} will be valid for the majority of ringed exoplanets in practice.

\revise{
It is possible to extend our postprocessing model to the case of an exoplanet with significant limb asymmetry.
Assuming that each morning and evening limb is homogeneous as in \citet{vonParis+16}, \citet{Macdonald+20}, and \citet{Espinoza&Jordan21} (see also top panel of Figure \ref{fig:limb_spectrum}), we can define the effective radius of each morning and evening terminator as
\begin{equation}\label{eq:spectrum_lim1}
    \pi R_{\rm mor/eve}^2=\pi R_{\rm 0}^2+\int_{\rm R_{\rm 0}}^{R_{\rm s}} 2\pi r[1-\exp(-\tau_{\rm mor/eve})]dr,
\end{equation}
where $\tau_{\rm mor/eve}$ is the chord optical depth at radial distance of $r$ in the morning/evening terminators.
The ring-free effective radius is then given by \citep{vonParis+16}
\begin{equation}
    R_{\rm ave}=\sqrt{ \frac{1}{2}(R_{\rm mor}^2+R_{\rm eve}^2)}.
\end{equation}
Application of Equation \eqref{eq:spectrum_post} to $R_{\rm ave}$ causes errors because the averaged radius no longer holds the information on the radius of each terminator.
One can instead respectively apply Equation \eqref{eq:spectrum_post} to each morning and evening terminator to derive the effective ringed radius of each terminator, such as
\begin{equation}\label{eq:spectrum_post_lim}
    \pi R_{\rm obs,mor/eve}^2=\pi R_{\rm mor/eve}^2 + S_{\rm ring,out}(R_{\rm mor/eve}) [1-\exp{(-\tau_{\rm ring,LOS})}].
\end{equation}
The observable radius is then given by
\begin{equation}\label{eq:Robs_lim}
    R_{\rm obs}=\sqrt{ \frac{1}{2}(R_{\rm obs,mor}^2+R_{\rm obs,eve}^2)}.
\end{equation}

\begin{figure}[t]
\centering
\includegraphics[clip, width=\hsize]{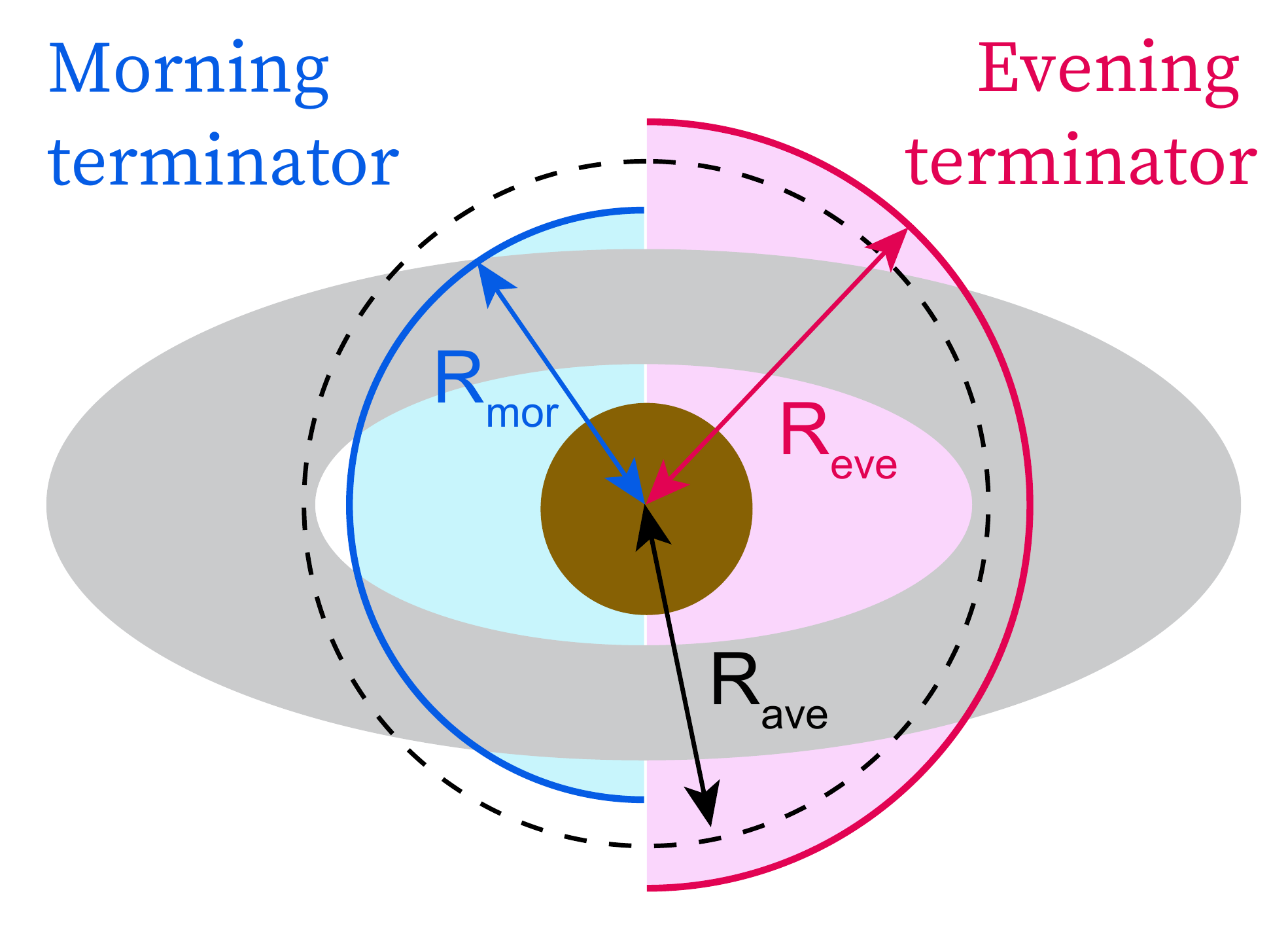}
\includegraphics[clip, width=\hsize]{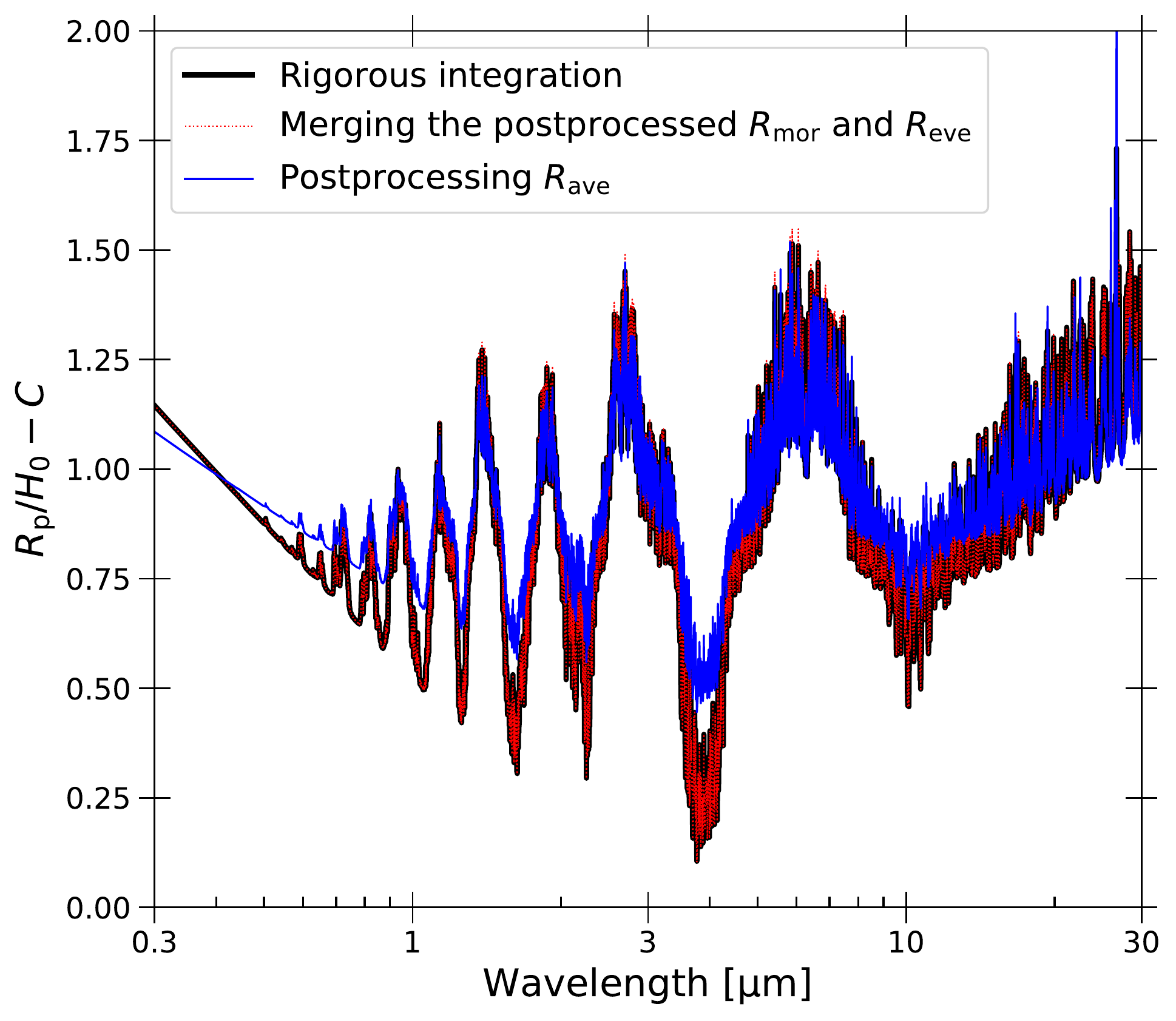}
\caption{\revise{(Top) Cartoon illustrating an exoplanet with significant limb asymmetry. The optically thick part of the morning and evening terminators have a radius of $R_{\rm mor}$ and $R_{\rm eve}$, respectively. The limb averaged radius $R_{\rm ave}$ is equivalent to the observable radius of a ring-free planet. (Bottom) Comparison of different methodologies for computing ringed transmission spectra of exoplanets with a limb asymmetry. The black line shows the spectrum computed by directly integrating the transmittance of each morning and evening terminator by splitting the ring-overlapped and ring-free areas, while the red line shows the spectrum computed by the modified postprocessing method of Equations \eqref{eq:spectrum_post_lim} and \eqref{eq:Robs_lim}.  The blue line shows the spectrum computed from the limb averaged radius $R_{\rm ave}$ by the postprocessing method of Section \ref{sec:approx}. 
We note that the black and red lines are almost completely superposed each other.
Each spectrum is normalized by the scale height at $r=R_{\rm 0}$ and $T=300~{\rm K}$.}
}
\label{fig:limb_spectrum}
\end{figure}
Our modified postprocessing method yields excellent agreement with the rigorous calculation.
The bottom panel of Figure \ref{fig:limb_spectrum} compares the spectrum computed by Equations \eqref{eq:spectrum_post_lim} and \eqref{eq:Robs_lim} with that computed by directly integrating the transmittance of each morning and evening annulus.
To compute the spectrum, we assume an isothermal atmosphere of $T=300~{\rm K}$ for the morning terminator and an isothermal cloudy atmosphere of $T=600~{\rm K}$ with a gray cloud top at $P={10}^{-5}~{\rm bar}$ for the evening terminator.

The direct application of the postprocessing method to the limb averaged radius $R_{\rm ave}$ yields noticeable discrepancies from the rigorous calculation.  In contrast, the modified morning and evening postprocessing method reproduces the direct integration quite well. 
Thus, one should apply Equations \eqref{eq:spectrum_post_lim} and \eqref{eq:Robs_lim} instead of Equation \eqref{eq:spectrum_post} when they model the exoplanet with limb asymmetry.
}


\section{Summary}\label{sec:summary}
In this study, we have presented a framework to characterize the atmospheric transmission spectra of exoplanets with circumplanetary rings.
We have established an analytical implementation of ring effects applicable for arbitrarily ring geometries (Section \ref{sec:method}).
We have also proposed a postprocessing method that can efficiently include the ring effects into precomputed ring-free spectra (Section \ref{sec:approx}).
Our postprocessing method has already been used to interpret the featureless transmission spectrum of HIP41378 f in a companion paper \citep{Alam+22}.
From a simple dynamical argument, we have also discussed that a thick ring may be sustainable only at cool equilibrium temperature ($\la$300~{\rm K}) as long as the ring's age is comparable to \revise{${\sim}3$ Gyr, which is} the typical system age of {\it Kepler} planets (\citealt{Berger+20}, Section \ref{sec:ring_surf}).

We have demonstrated that ringed planets show nearly flat transmission spectra for a wide range of viewing geometries, except for a nearly edge-on ring (Section \ref{sec:results}).
The featureless spectra can be consistent with recent observations for several extremely low-density exoplanets \citep[][]{Kreidberg+18,Libby-Roberts+20,Chachan+20,Alam+22}.
We have also found that the silicate features of a rocky ring, accessible by \emph{JWST} MIRI, may be observable when the ring's optical depth is around unity (Section \ref{sec:nongray}).
Although the detection of a ring's spectral features is viable only under limited conditions, if detected, this would provide tight constrains on the physical and compositional properties of exoplanetary rings, which in turn offer clues to the origin of exoplanetary ring systems.

\acknowledgments 
We thank the anonymous referee for providing helpful comments to improve this paper.
We also thank Munazza Alam, Peter Gao, and Yuki A. Tanaka for motivating this project.
K.O. was supported by a JSPS Overseas Research Fellowship.  J.J.F. is supported by an award from the Simons Foundation.


\end{document}